\documentclass[
  aps,                  
  prd,                  
  reprint,              
  nofootinbib,          
  superscriptaddress,   
  showpacs,             
  showkeys,             
  longbibliography      
]{revtex4-1}

\usepackage[utf8]{inputenc}
\usepackage{amsfonts,amsmath,amssymb}
\usepackage{bm,dsfont} 
\usepackage{url}
\usepackage{graphicx}
\usepackage{hyperref}
\usepackage{multirow}

\newcommand{\nwse}[3]{\ensuremath{#1^{#2}_{\phantom{#2} #3}}}

\newcommand{\bring}[1]{\mathring{\bar{#1}}}
\newcommand{\hodge}{{\ast}} 

\newcommand{\udec}{Departamento de Fí­sica, Universidad de Concepción, Casilla 160-C, Concepción, Chile}
\newcommand{\unal}{Departamento de Fí­­sica, Universidad Nacional de Colombia, Bogotá, Colombia}
\newcommand{\unap}{Facultad de Ingeniería y Arquitectura, Universidad Arturo Prat, Iquique, Chile}

\begin{document}

\title{Linear and Second-order Geometry Perturbations on Spacetimes with Torsion}

\affiliation{\unap}

\author{Fernando Izaurieta}
\email{fizaurie@udec.cl}
\affiliation{\udec}

\author{Eduardo Rodríguez}
\email{eduarodriguezsal@unal.edu.co}
\affiliation{\unal}

\author{Omar Valdivia}
\email{ovaldivi@unap.cl}
\affiliation{\unap}

\date{\today}

\begin{abstract}
In order to study gravitational waves in any realistic astrophysical scenario,
one must consider geometry perturbations up to second order.
Here, we present a general technique for studying linear and
quadratic perturbations on a spacetime with torsion.
Besides the standard metric mode, a ``torsionon'' perturbation mode appears.
This torsional mode will be able to propagate only in a certain kind of theories.
\end{abstract}

\pacs{04.50.+h}

\keywords{Nonvanishing Torsion, Gravitational Waves, Cartan Geometry.}

\maketitle

%
%


\section{Introduction}

Besides opening up the new field of gravitational-wave astronomy,
gravitational waves have proven to be a valuable tool
in order to constraint alternative theories of gravity.
In particular, the detection of GW170817~\cite{TheLIGOScientific:2017qsa}
refuted whole families of theories~\cite{Ezquiaga:2018btd,Ezquiaga:2017ekz,Sakstein:2017xjx}.
Together with other experimental results~\cite{Mukherjee:2017fqz},
this amounts to a renewed challenge on alternative gravitational theories.

The standard approach to gravitational waves is based on Riemannian geometry.
We start by considering a metric $\bar{g}_{\mu\nu}$
composed of a background metric $g_{\mu \nu}$
and a small perturbation $h_{\mu\nu}$ on it,
\begin{equation}
  g_{\mu\nu} \mapsto \bar{g}_{\mu\nu} =
  g_{\mu\nu} + h_{\mu\nu},
  \label{Ec_Metric_Perturb}
\end{equation}
and from this metric we calculate the Christoffel connection,
the Riemann curvature, the behavior of the field equations
under the perturbations, etc.

To repeat this success for theories formulated using Rie\-mann--Cartan (RC) geometry
instead of Riemannian geometry may seem difficult.
The basic premise of RC geometry is
to consider the connection $\Gamma_{\mu\nu}^{\lambda}$
as a new degree of freedom.
However, the connection is still metric-compatible, i.e., partly metric-dependent.
Thus, the torsion of the connection,
\begin{equation}
  \nwse{T}{\lambda}{\mu\nu} =
  \Gamma_{\mu\nu}^{\lambda} - \Gamma_{\nu\mu}^{\lambda},
\end{equation}
incorporates the 24 new independent degrees of freedom of a RC-space in 4~dimensions.

The simplest theory formulated using RC geometry
is Einstein--Cartan--Sciama--Kibble (ECSK)
gravity~\cite{Kib61,Sciama:1964wt,RevModPhys.48.393},
but many other theories in four and higher dimensions
are naturally expressed in this framework
(see, e.g., Refs.~\cite{Hehl:1994ue,Mardones:1990qc,BlagoHehlGTG2013}).

The presence of propagating torsion modes in this class of theories
has been the subject of much investigation; for a recent assessment, see
Refs.~\cite{Boos:2016cey,Blagojevic:2017ssv,Blagojevic:2018dpz,Karananas:2014pxa,*Karananas:2015}
and references therein.

Traditionally, theoretical work on gravitational waves, whether including torsion or not, has used exclusively the tensor formalism, focusing on the metric and (if torsion is a concern) the affine connection perturbations. It is the purpose of this article to show that Cartan's elegant and powerful exterior calculus, which, when applied to gravitational theory, treats the vielbein and spin connection one-forms as independent variables (see, e.g., Ref.~\citep{Zanelli:2005sa}), is also up to the task. Crucial in accomplishing this goal is the introduction of certain operators whose action on differential $p$-forms mimics tensor operations without ever writing spacetime indices explicitly. In this paper, we use Cartan's formalism to parametrize linear and second-order perturbations on the vielbein and the spin connection in a manner that clearly separates the metric from the affine degrees of freedom, while remaining well-suited for practical computations.

As part of an earlier work, in Ref.~\cite{Valdivia:2017sat}
we presented a quick introduction on how this can be accomplished to linear order.
In this paper, we will show in detail
how it is done up to second order in perturbations.
Going to second order is essential
in order to construct realistic astrophysical models.
As explained in Chapter~1.4 of Ref.~\cite{Maggiore:1900zz},
we have to be able to split the field equations
as a low-frequency background, $g_{\mu\nu}$,
plus a high-frequency term, $h_{\mu\nu}$.
Since the product of terms of similar high frequencies
can give rise to low-frequency effects,
we have to consider linear and quadratic terms in $h_{\mu\nu}$
when constructing $\bar{g}^{\mu\nu}$,
$\bar{\Gamma}_{\mu\nu}^{\lambda}$,
$\nwse{\bar{R}}{\rho}{\sigma\mu\nu}$, etc.,
in the context of Riemannian geometry.
This means that second-order terms become essential to define
an effective stress-energy tensor for the perturbations.

The current article deals only with \emph{kinematics}.
We will not construct any model, and we will not discuss any dynamics.
The only purpose of this article is to provide a practical tool
for anyone interested in studying geometry perturbations
on RC geometry with nonvanishing torsion.
Our results hold for any theory based on RC geometry.

Besides the graviton, there is a new mode, the ``torsion\-on.''%
\footnote{The word ``torsionon,'' which rolls off the tongue in Spanish, may be a bit hard to pronounce in English. J.~Boos and F.~W.~Hehl have called these degrees of freedom ``gravitational W and Z gauge bosons''~\cite{Boos:2016cey}, while F.~W.~Hehl had earlier suggested the name ``roton''~\cite{Hehl1980}.}
Whether or not it propagates depends on the choice of Lagrangian.
For instance, in standard ECSK gravity and other similar theories~\cite{Shimada:2018lnm},
this new mode will not propagate in a vacuum.
However, in nonminimally coupled theories
(see, e.g., Ref.~\cite{Valdivia:2017sat,Bonder:2018mfz})
or the MacDowell--Mansouri theory \cite{PhysRevLett.38.739,Wise:2006sm},
it will.


\section{Geometry Perturbations on a Spacetime with Torsion}

\subsection{First-order Formalism}
\label{sec:1storder}

Let $M$ be a spacetime manifold and let
$g_{\mu\nu}$ be the
coordinate components of its metric tensor at some point $P$.
The orthonormal coframe $\nwse{e}{a}{\mu}$ is implicitly defined through the relation
\begin{equation}
  g_{\mu\nu} = \eta_{ab} \nwse{e}{a}{\mu} \nwse{e}{b}{\nu},
  \label{eq:g=hee}
\end{equation}
where $\eta_{ab}$ is the Minkowski metric.
We reserve the term \emph{vielbein} for the collection of one-forms given by
$e^{a} = \nwse{e}{a}{\mu} \mathrm{d} x^{\mu}$.
Since knowledge of the vielbein at every spacetime point is equivalent
to knowing the metric at every spacetime point, one can
shift focus from $g_{\mu\nu}$ to $e^{a}$ when studying the spacetime geometry.
The use of differential forms guarantees that the vielbein
remains invariant under general coordinate transformations.

A crucial stage in the development of the first-order formalism comes with
the realization that the orthonormal co\-frame at $P$ is not unique.
In terms of the vielbein, this means that the rotated vielbein
$e'^{a} = \nwse{\Lambda}{a}{b} e^{b}$
is associated with the same spacetime metric $g_{\mu\nu}$ as $e^{a}$
[via eq.~(\ref{eq:g=hee})]
provided that $\nwse{\Lambda}{a}{b}$ represents a
Lorentz transformation, satisfying
$\eta_{ab} = \nwse{\Lambda}{c}{a} \nwse{\Lambda}{d}{b} \eta_{cd}$.
Such a local Lorentz rotation amounts to a gauge transformation
on $e^{a}$, with the vielbein behaving as a Lorentz vector (of one-forms).

The exterior derivative of the vielbein,
$\mathrm{d} e^{a}$, is a two-form, and, consequently, invariant under
general coordinate transformations.
The fact that Lorentz gauge transformations are local implies, however,
that $\mathrm{d} e^{a}$ does not inherit the vector nature of $e^{a}$.
To deal with this problem, we introduce the \emph{spin connection}%
\footnote{A more appropriate name would be \emph{Lorentz connection}.
The ``spin connection'' name is often used in the physics literature
because $\omega^{ab}$ is needed to take the covariant derivative
of spinor fields.}
$\omega^{ab}$
and define the Lorentz-covariant exterior derivative of the vielbein
(and similarly for any Lorentz vector form) as
\begin{equation}
  \mathrm{D} e^{a} = \mathrm{d} e^{a} + \nwse{\omega}{a}{b} \wedge e^{b},
\end{equation}
where $\mathrm{d} + \omega$ transforms as a Lorentz tensor.
The covariant exterior derivative of the vielbein defines
the \emph{torsion} two-form, $T^{a} = \mathrm{D} e^{a}$.

Unlike the ordinary exterior derivative, which satisfies $\mathrm{d}^{2} = 0$,
the covariant exterior derivative can yield nonzero results when applied
repeatedly. A direct calculation shows that the covariant exterior derivative
of the torsion can be written as
$\mathrm{D} T^{a} = \nwse{R}{a}{b} \wedge e^{b}$,
where
\begin{equation}
  \nwse{R}{a}{b} = \mathrm{d} \nwse{\omega}{a}{b} +
  \nwse{\omega}{a}{c} \wedge \nwse{\omega}{c}{b}
  \label{eq:Rab}
\end{equation}
is a Lorentz-tensor two-form called \emph{Lorentz curvature}.
Despite being defined in terms of the non-tensorial spin connection,
the Lorentz curvature transforms as a tensor under local Lorentz transformations.

No new objects appear by further application of the covariant exterior derivative,
since (as can be easily shown) $\mathrm{D} R^{ab} = 0$.

The spin connection represents a new degree of freedom in the theory,
independent from the metric degrees of freedom encoded in $e^{a}$.
Its tensor-formalism analog is the \emph{affine} connection $\Gamma^{\lambda}_{\mu\nu}$.
A direct link between the two can be established by means of the so-called
``vielbein postulate,''
\begin{equation}
  \partial_{\mu} \nwse{e}{a}{\nu} +
  \nwse{\omega}{a}{b\mu} \nwse{e}{b}{\nu} - 
  \Gamma^{\lambda}_{\mu\nu} \nwse{e}{a}{\lambda} = 0.
\end{equation}

Just like the affine connection, the spin connection can be split into
a torsionless piece $\mathring{\omega}^{ab}$, which satisfies
\begin{equation}
  \mathrm{d} e^{a} + \nwse{\mathring{\omega}}{a}{b} \wedge e^{b} = 0,
  \label{eq:T=0}
\end{equation}
and a tensor one-form $\kappa^{ab}$ called the \emph{contorsion}.
The torsionless piece is uniquely determined [via eq.~(\ref{eq:T=0})]
in terms of the vielbein and its derivatives,
meaning that the contorsion, defined as the difference
$\kappa^{ab} = \omega^{ab} - \mathring{\omega}^{ab}$,
carries all affine degrees of freedom within itself.
Torsion and contorsion are related by
$T^{a} = \nwse{\kappa}{a}{b} \wedge e^{b}$.

The fact that the torsionless connection $\mathring{\omega}^{ab}$
is indeed a connection means that a curvature tensor and a covariant derivative
can be associated with it.
We define the \emph{Riemann curvature} two-form $\mathring{R}^{ab}$
as the purely metric concoction [cf.~eq.~(\ref{eq:Rab})]
\begin{equation}
  \nwse{\mathring{R}}{a}{b} = \mathrm{d} \nwse{\mathring{\omega}}{a}{b} +
  \nwse{\mathring{\omega}}{a}{c} \wedge \nwse{\mathring{\omega}}{c}{b}.
\end{equation}
The Lorentz and Riemann curvatures are related by
\begin{equation}
  R^{ab} = \mathring{R}^{ab} + \mathring{\mathrm{D}} \kappa^{ab} +
  \nwse{\kappa}{a}{c} \wedge \nwse{\kappa}{c}{b},
\end{equation}
where $\mathring{\mathrm{D}}$ stands for the Lorentz-covariant derivative
computed with the torsionless connection $\mathring{\omega}^{ab}$.

With two connections, two curvatures,
and two covariant exterior derivatives,
things may get cumbersome.
A large part of what follows is an attempt to keep the complications at bay
by carefully parametrizing perturbations on the geometry in such a way as to
easily track the metric and affine degrees of freedom involved,
while keeping calculations as simple as possible.


\subsection{Default Parametrization of Perturbations}
\label{sec:naive}

Let the pair $\left( e^{a}, \omega^{ab} \right)$
describe a background geometry as independent degrees of freedom.
On this background spacetime we place independent perturbations
$H^{a}$ and $u^{ab}$ such that the new, perturbed geometry can be described
by the pair $\left( \bar{e}^{a}, \bar{\omega}^{ab} \right)$, with
\begin{align}
  e^{a} & \mapsto \bar{e}^{a} =
  e^{a} + \frac{1}{2}H^{a},
  \label{Ec_vielbein_Perturb} \\
  \omega^{ab} & \mapsto \bar{\omega}^{ab} =
  \omega^{ab} + u^{ab}.
  \label{Ec_Spin_Connection_Perturb}
\end{align}

The vielbein perturbation, being a one-form, can be written as
$H^{a} = \nwse{H}{a}{b} e^{b}$, where $H^{ab}$ is a zero-form.
The antisymmetric part of $H^{ab}$ corresponds to an infinitesimal local
Lorentz transformation on $e^{a}$, which does not change the spacetime metric.
This means that, without any loss of generality, we can restrict
$H^{ab}$ to be symmetric, $H^{ba} = H^{ab}$.
(See Appendix~\ref{sec:Lorentz} for further details on why this can be done).

Let us write the perturbation on the spacetime metric as
\begin{equation}
  g_{\mu\nu} \mapsto \bar{g}_{\mu\nu} =
  g_{\mu\nu} + h_{\mu\nu}.
\end{equation}
If $g_{\mu\nu}$ and $\bar{g}_{\mu\nu}$ are each related to
$e^{a}$ and $\bar{e}^{a}$ through eq.~(\ref{eq:g=hee}),
then the metric and the vielbein perturbations are related by
\begin{equation}
  h_{\mu\nu} = H_{\mu \nu} +
  \frac{1}{4} H_{\lambda\mu} \nwse{H}{\lambda}{\nu},
\end{equation}
where $H_{\mu\nu} = H_{ab} \nwse{e}{a}{\mu} \nwse{e}{b}{\nu}$.
The inverse relation can be written as the series
\begin{equation}
  \nwse{H}{a}{\mu} = \nwse{h}{a}{\mu} -
  \frac{1}{4} \nwse{h}{a}{\rho} \nwse{h}{\rho}{\mu} +
  \frac{1}{8} \nwse{h}{a}{\lambda} \nwse{h}{\lambda}{\rho} \nwse{h}{\rho}{\mu} +
  \cdots.
  \label{eq:Hh}
\end{equation}
In particular, this means that even if $h_{\mu\nu}$ and $H_{\mu\nu}$
differ at the quadratic level, they are but two equivalent ways
of parametrizing the same (metric) degrees of freedom.

Having established the relation between the metric and the vielbein
perturbations, the following course of action presents itself to us.
Let $L = L \left( e^{a}, \omega^{ab} \right)$
be the four-form Lagrangian density which defines the theory,
so that its field equations can be written as
$\delta L / \delta e^{a} = 0$ and
$\delta L / \delta \omega^{ab} = 0$.
By replacing eqs.~(\ref{Ec_vielbein_Perturb})--(\ref{Ec_Spin_Connection_Perturb})
into the field equations, one might attempt to solve
$\delta L / \delta \omega^{ab} = 0$ for $u^{ab}$
and replace the result in $\delta L / \delta e^{a} = 0$,
thus eliminating $u^{ab}$ from consideration.
Experience shows this course of action to be
an algebraic nightmare even in simple cases.

A much more practical way to proceed goes as follows.
Let us start by splitting the background and the perturbed spin connections
into the corresponding torsionless connections plus contorsion tensors
(see section~\ref{sec:1storder}),
\begin{align}
  \omega^{ab} & = \mathring{\omega}^{ab} + \kappa^{ab}, \\
  \bar{\omega}^{ab} & = \bring{\omega}^{ab} + \bar{\kappa}^{ab}.
\end{align}
Define now
\begin{align}
  \mathring{u}^{ab} & = \bring{\omega}^{ab} - \mathring{\omega}^{ab}, \\
  q^{ab} & = \bar{\kappa}^{ab} - \kappa^{ab}.
\end{align}
Here, $\mathring{u}^{ab}$ represents the perturbation on the torsionless
piece of the spin connection. This perturbation is not an independent
degree of freedom, but rather can be entirely written in terms of the
vielbein perturbation, $H^{a}$, and its derivatives.
On the other hand, $q^{ab}$ encodes the perturbation on the contorsion,
which is fully independent from the vielbein perturbation.
The full spin connection perturbation, $u^{ab}$,
is the sum of these two contributions:
\begin{equation}
  u^{ab} = \mathring{u}^{ab} + q^{ab}.
\end{equation}

To be consistent, $\bring{\omega}^{ab}$ must be a torsionless connection
for the perturbed geometry $\left( \bar{e}^{a}, \bar{\omega}^{ab} \right)$.
The torsionless condition
$\mathrm{d} \bar{e}^{a} + \nwse{\bring{\omega}}{a}{b} \wedge \bar{e}^{b} = 0$
imposes the following constraint on $\mathring{u}^{ab}$:
\begin{equation}
  \frac{1}{2} \mathring{\mathrm{D}} H^{a} +
  \nwse{\mathring{u}}{a}{b} \wedge \left(
    e^{b} + \frac{1}{2} H^{b}
  \right) = 0.
  \label{Ec_deduc_u_torsionless}
\end{equation}

It is possible to solve eq.~(\ref{Ec_deduc_u_torsionless})
for $\mathring{u}^{ab}$ as a series in $H^{a}$,
\begin{equation}
  \mathring{u}_{ab} =
  \mathring{u}_{ab}^{\left( 1 \right)} +
  \mathring{u}_{ab}^{\left( 2 \right)} +
  \mathcal{O} \left( H^{3} \right),
\end{equation}
where the linear and quadratic terms are explicitly given by
\begin{align}
  \mathring{u}_{ab}^{\left( 1 \right)} & = -
  \frac{1}{2} \left(
    \mathrm{I}_{a} \mathring{\mathrm{D}} H_{b} -
    \mathrm{I}_{b} \mathring{\mathrm{D}} H_{a}
  \right),
  \label{Ec_u_ab(1)} \\
  \mathring{u}_{ab}^{\left( 2 \right)} & =
  \frac{1}{8} \mathrm{I}_{ab} \left(
    \mathrm{\mathring{D}}H_{c}\wedge H^{c}
  \right) +
  \nonumber \\ & +
  \frac{1}{2} \left[
    \mathrm{I}_{b} \left(
      \mathring{u}_{ac}^{\left( 1 \right)} \wedge H^{c}
    \right)  -
    \mathrm{I}_{a} \left(
      \mathring{u}_{bc}^{\left( 1 \right)} \wedge H^{c}
    \right)
  \right].
  \label{Ec_u_ab(2)}
\end{align}
The $\mathrm{I}^{a}$ and $\mathrm{I}^{ab}$ operators
in eqs.~(\ref{Ec_u_ab(1)})--(\ref{Ec_u_ab(2)}) are given by
$\mathrm{I}^{a} = -\hodge \left( e^{a} \wedge \hodge \right.$ and
$\mathrm{I}^{ab} = \hodge \left( e^{a} \wedge e^{b} \wedge \hodge \right.$
on a four-dimensional spacetime with Lorentzian signature,
where $\hodge$ is the Hodge dual.
See Appendix~\ref{sec:Iop} for their formal definitions
on a $d$-dimensional manifold with arbitrary signature.

This is certainly progress; we have managed to split the spin connection
perturbation into a torsionless piece $\mathring{u}^{ab}$ plus an
independent contorsion perturbation $q^{ab}$, and to cast $\mathring{u}^{ab}$
as a series in $H^{a}$, with the first two terms given by
eqs.~(\ref{Ec_u_ab(1)})--(\ref{Ec_u_ab(2)}).
Unfortunately, this is not yet good enough.
To see why, note that eqs.~(\ref{Ec_u_ab(1)})--(\ref{Ec_u_ab(2)}) give us
$\mathring{u}^{ab}$ in terms of the torsionless covariant derivative
$\mathring{\mathrm{D}} = \mathrm{d} + \mathring{\omega}$.
The field equations, however, use $\mathrm{D} = \mathrm{d} + \omega$
and the Lorentz curvature $R^{ab}$ instead of the Riemann curvature
$\mathring{R}^{ab}$.
Mixing both operators and curvatures proves to be a recipe for algebraic disaster.


\subsection{Useful Parametrization of Perturbations}
\label{sec:artful}

In order to deal with the derivative-mixing problem,
we define the new variables
\begin{align}
  U^{ab} & = \mathring{u}^{ab} - \Delta^{ab}, \\
  V^{ab} & = q^{ab} + \Delta^{ab},
\end{align}
where $\Delta^{ab}$ is an antisymmetric Lorentz-tensor one-form to be determined.
See Table~\ref{tab:summary} for a quick reminder of all variables defined so far.
Clearly, the spin connection perturbation can be written as
\begin{equation}
  u^{ab} = \mathring{u}^{ab} + q^{ab} = U^{ab} + V^{ab},
\end{equation}
so $U^{ab}$ and $V^{ab}$ are to be regarded as a different choice
for the metric-dependent and metric-independent contributions
to $u^{ab}$, respectively.


\begin{table}
  \begin{ruledtabular}
  \begin{tabular}{lccc}
    & \multirow{2}{*}{Background} & \multicolumn{2}{c}{Perturbation} \\
    \cline{3-4}
    &  & Default & Useful \\
    \hline
    Independent & $\kappa^{ab}$            & $q^{ab}$            & $V^{ab}$ \\
    Derived     & $\mathring{\omega}^{ab}$ & $\mathring{u}^{ab}$ & $U^{ab}$ \\
    Mixed (sum) & $\omega^{ab}$            & $u^{ab}$            & $u^{ab}$ \\
  \end{tabular}
  \end{ruledtabular}
  \caption{\label{tab:summary}
  Torsion is most naturally handled through the first-order formalism.
  The spin connection can be split as
  $\omega^{ab} = \mathring{\omega}^{ab} + \kappa^{ab}$,
  where the torsionless, Riemannian connection $\mathring{\omega}^{ab}$
  is derived from the vielbein $e^{a}$ (and its derivatives) via
  $\mathrm{d} e^{a} + \protect \nwse{\mathring{\omega}}{a}{b} \wedge e^{b} = 0$,
  and the \emph{contorsion} $\kappa^{ab}$ represents a new,
  independent degree of freedom.
  (The two-form torsion is given by
  $T^{a} = \protect \nwse{\kappa}{a}{b} \wedge e^{b}$).
  When adding a perturbation $u^{ab}$ to a background spin connection,
  the default approach consists of also splitting $u^{ab}$ into a torsionless
  piece $\mathring{u}^{ab}$
  (completely determined from the vielbein perturbation, $H^{a}$)
  and an independent piece, $q^{ab}$.
  As we argue in the text, it proves far more convenient to artfully
  modify this splitting as $u^{ab} = U^{ab} + V^{ab}$, where
  $U^{ab} = \mathring{u}^{ab} - \Delta^{ab}$ is,
  like $\mathring{u}^{ab}$ itself,
  completely determined from $H^{a}$ (and its derivatives),
  and $V^{ab} = q^{ab} + \Delta^{ab}$
  is to be regarded as an independent perturbation.
  A convenient choice for $\Delta^{ab}$ is
  then required to satisfy eq.~(\ref{eq:Delta-constraint}).}
\end{table}


In terms of $U^{ab}$ and $\Delta^{ab}$,
the $\mathring{u}$-constraint~(\ref{Ec_deduc_u_torsionless})
now takes the form
\begin{align}
  \lefteqn{
    \frac{1}{2} \mathrm{D} H^{a} +
    \nwse{U}{a}{b} \wedge \left( e^{b} + \frac{1}{2} H^{b} \right) +
  } \nonumber \\ & +
  \nwse{\Delta}{a}{b} \wedge \left( e^{b} + \frac{1}{2} H^{b} \right) - 
  \frac{1}{2} \nwse{\kappa}{a}{b} \wedge H^{b} = 0.
  \label{Ec_U_definition}
\end{align}
A long, hard look at eqs.~(\ref{Ec_u_ab(1)})--(\ref{Ec_u_ab(2)})
suggests that it may be wise to demand $\Delta_{ab}$ to satisfy
\begin{align}
  \lefteqn{
    \Delta_{ab} \wedge \left( e^{b} + \frac{1}{2} H^{b} \right) -
    \frac{1}{2} \kappa_{ab} \wedge H^{b} =
  } \nonumber \\ & =
  \frac{1}{2} \mathrm{I}_{a} \left[
    H^{b} \wedge T_{b} - \frac{1}{2} H^{b} \wedge \mathrm{I}_{b} \left(
      H^{c} \wedge T_{c}
    \right)
  \right] +
  \mathcal{O} \left( H^{3} \right).
  \label{eq:Delta-constraint}
\end{align}
Crucially, we are now able to solve eq.~(\ref{eq:Delta-constraint})
for $\Delta_{ab}$ as a series in $H^{a}$,
\begin{equation}
  \Delta_{ab} =
  \Delta_{ab}^{\left( 1 \right)} +
  \Delta_{ab}^{\left( 2 \right)} +
  \mathcal{O} \left( H^{3} \right),
\end{equation}
where the linear and quadratic terms are given explicitly by
\begin{align}
  \Delta_{ab}^{\left( 1 \right)} & =
  \frac{1}{2} \left[
    \mathrm{I}_{a} \left( \kappa_{bc} \wedge H^{c} \right) -
    \mathrm{I}_{b} \left( \kappa_{ac} \wedge H^{c} \right)
  \right],
  \\
  \Delta_{ab}^{\left( 2 \right)} & = -
  \frac{1}{8} \mathrm{I}_{ab} \left(
    H^{c} \wedge \kappa_{cd} \wedge H^{d}
  \right) +
  \nonumber \\ & -
  \frac{1}{2} \left[
    \mathrm{I}_{a} \left( \Delta_{bc}^{\left( 1 \right)} \wedge H^{c} \right) -
    \mathrm{I}_{b} \left( \Delta_{ac}^{\left( 1 \right)} \wedge H^{c} \right)
  \right].
\end{align}
With this choice for $\Delta_{ab}$, eq.~(\ref{Ec_U_definition})
gives us a much nicer version of eqs.~(\ref{Ec_u_ab(1)})--(\ref{Ec_u_ab(2)}).
Writing
\begin{equation}
  U_{ab} =
  U_{ab}^{\left( 1 \right)} +
  U_{ab}^{\left( 2 \right)} +
  \mathcal{O} \left( H^{3} \right),
  \label{eq:Useries}
\end{equation}
we now find
[cf.~eqs.~(\ref{Ec_u_ab(1)})--(\ref{Ec_u_ab(2)})]
\begin{align}
  U_{ab}^{\left( 1 \right)} & = -\frac{1}{2} \left(
    \mathrm{I}_{a} \mathrm{D} H_{b} -
    \mathrm{I}_{b} \mathrm{D} H_{a}
  \right),
  \label{eq:U1}
  \\
  U_{ab}^{\left( 2 \right)} & =
  \frac{1}{8} \mathrm{I}_{ab} \left(
    \mathrm{D} H_{c} \wedge H^{c}
  \right) +
  \nonumber \\ &
  - \frac{1}{2} \left[
    \mathrm{I}_{a} \left( U_{bc}^{\left( 1 \right)} \wedge H^{c} \right) -
    \mathrm{I}_{b} \left( U_{ac}^{\left( 1 \right)} \wedge H^{c} \right)
  \right].
  \label{eq:U2}
\end{align}
Eqs.~(\ref{eq:U1})--(\ref{eq:U2}) allow us to compute $U^{ab}$
up to quadratic terms in $H^{a}$ using the same Lorentz-covariant
exterior derivative $\mathrm{D}$ that appears in the field equations.

The perturbed torsion and curvature can now be written in terms of the
new variables $U^{ab}$ and $V^{ab}$. For the torsion we find
\begin{align}
  T_{a} \mapsto \bar{T}_{a} & = T_{a} +
  V_{ab} \wedge \left( e^{b} + \frac{1}{2} H^{b} \right) +
  \nonumber \\ & -
  \frac{1}{2} \mathrm{I}_{a} \left[
    H^{b} \wedge T_{b} -
    \frac{1}{2} H^{b} \wedge \mathrm{I}_{b}
    \left( H^{c} \wedge T_{c} \right)
  \right].
  \label{Eq_perturb_Torsion}
\end{align}
For the curvature we write
\begin{equation}
  R^{ab} \mapsto \bar{R}^{ab} = R^{ab} +
  R_{\left( 1 \right)}^{ab} +
  R_{\left( 2 \right)}^{ab} +
  \mathcal{O} \left( H^{3} \right),
  \label{eq:Rseries}
\end{equation}
where the linear and quadratic terms are given by
\begin{align}
  R_{\left( 1 \right)}^{ab} & =
  \mathrm{D} U_{\left( 1 \right)}^{ab} + \mathrm{D} V^{ab},
  \label{Eq_R1}
  \\
  R_{\left( 2 \right)}^{ab} & =
  \mathrm{D} U_{\left( 2 \right)}^{ab} +
  \left( U^{a}_{\left( 1 \right) c} + \nwse{V}{a}{c} \right) \wedge
  \left( U_{\left( 1 \right)}^{cb} + V^{cb} \right).
  \label{Eq_R2}
\end{align}

These expressions reduce to the standard perturbation case
[see, e.g., Ref.~\cite{Maggiore:1900zz}, eqs.~(1.113)--(1.114)]
in the torsionless limit.


\section{Final Remarks}

In this paper, we have shown that in the context of Riemann--Cartan geometry
perturbations are best described as
\begin{align}
  e^{a} & \mapsto \bar{e}^{a} =
  e^{a} + \frac{1}{2} H^{a},
  \\
  \omega^{ab} & \mapsto \bar{\omega}^{ab} =
  \omega^{ab} + U^{ab} \left( H, \partial H \right) + V^{ab}.
\end{align}
Here, $H^{a}$ is related to the standard metric perturbation by
[cf.~eq.~(\ref{eq:Hh})]
\begin{equation}
  \nwse{H}{a}{\mu} = \nwse{h}{a}{\mu} -
  \frac{1}{4} \nwse{h}{a}{\rho} \nwse{h}{\rho}{\mu} +
  \frac{1}{8} \nwse{h}{a}{\lambda} \nwse{h}{\lambda}{\rho} \nwse{h}{\rho}{\mu} +
  \cdots.
\end{equation}
The $H$-dependent contribution to the spin connection perturbation is given by
[cf.~eqs.~(\ref{eq:Useries})--(\ref{eq:U2})]
\begin{align}
  U_{ab} & = -\frac{1}{2} \left(
    \mathrm{I}_{a} \mathrm{D} H_{b} -
    \mathrm{I}_{b} \mathrm{D} H_{a}
  \right) + 
  \frac{1}{8} \mathrm{I}_{ab} \left(
    \mathrm{D} H_{c} \wedge H^{c}
  \right) +
  \nonumber \\ & + \frac{1}{4}
    \mathrm{I}_{a} \left[
      \left(
        \mathrm{I}_{b} \mathrm{D} H_{c} - \mathrm{I}_{c} \mathrm{D} H_{b}
      \right)
      \wedge H^{c}
    \right] +
  \nonumber \\ & - \frac{1}{4}
    \mathrm{I}_{b} \left[
      \left(
        \mathrm{I}_{a} \mathrm{D} H_{c} - \mathrm{I}_{c} \mathrm{D} H_{a}
      \right)
      \wedge H^{c}
    \right] +
  \mathcal{O} \left( H^{3} \right).
\end{align}
Finally, $V^{ab}$ is an $H$-independent torsional mode.
See Appendix~\ref{sec:Iop} for the definition of the $\mathrm{I}$-operators.

In terms of this parametrization of the geometry perturbation,
torsion and curvature can be written as
[cf.~eq.~(\ref{Eq_perturb_Torsion})]
\begin{align}
  T_{a} \mapsto \bar{T}_{a} & = T_{a} +
  V_{ab} \wedge \left( e^{b} + \frac{1}{2} H^{b} \right) +
  \nonumber \\ & -
  \frac{1}{2} \mathrm{I}_{a} \left[
    H^{b} \wedge T_{b} -
    \frac{1}{2} H^{b} \wedge \mathrm{I}_{b}
    \left( H^{c} \wedge T_{c} \right)
  \right],
\end{align}
and
[cf.~eqs.~(\ref{eq:Rseries})--(\ref{Eq_R2})]
\begin{align}
  R^{ab} \mapsto \bar{R}^{ab} & = R^{ab} +
  \mathrm{D} U_{\left( 1 \right)}^{ab} + \mathrm{D} V^{ab} +
  \mathrm{D} U_{\left( 2 \right)}^{ab} +
  \nonumber \\ & +
  \left( U^{a}_{\left( 1 \right) c} + \nwse{V}{a}{c} \right) \wedge
  \left( U_{\left( 1 \right)}^{cb} + V^{cb} \right) +
  \mathcal{O} \left( H^{3} \right).  
\end{align}

Our analysis has remained kinematic throughout.
The actual evolution of perturbations will depend, of course,
on the underlying dynamical theory.
Note, however, that our results imply that there will be significant differences
in the propagation of the graviton, $H^{a}$, versus the ``torsionon,'' $V^{ab}$.
In eqs.~(\ref{Eq_R1})--(\ref{Eq_R2}), the
$\mathrm{D} U_{\left( 1 \right)}^{ab}$ and
$\mathrm{D} U_{\left( 2 \right)}^{ab}$ terms
produce second-order derivatives of $H^{a}$
(as it must be in order to have gravitational waves).
The case of $V^{ab}$ is a bit different.
The perturbation of torsion
[cf.~eq.~(\ref{Eq_perturb_Torsion})]
does not include derivatives of $V^{ab}$.
The only place where these derivatives appear
is in $R_{\left( 1 \right)}^{ab}$ [cf.~eq.~(\ref{Eq_R1})].
This means that in a theory where the field equations
do not include derivatives of $R^{ab}$
(e.g., Lovelock or Horndeski),
the propagation equations for the torsionon will include at most
first-order derivatives of $V^{ab}$.
In order to have a wave equation for $V^{ab}$,
the field equations must include terms such as
$\mathrm{D} \, \hodge R^{ab}$,
where $\hodge$ stands for the Hodge dual (see Appendix~\ref{sec:Iop}).


\begin{acknowledgments}
We are grateful to
José Barrientos, Jens Boos, Oscar Castillo-Felisola, Fabrizio Cordonier-Tello, Cristóbal Corral, Nicolás González, Perla Medina, Daniela Narbona, Julio Oliva, Francisca Ramírez, Patricio Salgado, Sebastián Salgado, Jorge Zanelli,
and Alfonso Zerwekh for many enlightening conversations.
FI acknowledges financial support from the Chilean government
through Fondecyt grants 1150719 and 1180681.
OV acknowledges VRIIP UNAP for financial support through Project VRIIP0258-18.
\end{acknowledgments}


\appendix

\section{The \texorpdfstring{$\mathrm{I}$}{I}-operators}
\label{sec:Iop}

Let $M$ be a $d$-dimensional differentiable manifold endowed with a metric
$g_{\mu\nu} = \eta_{ab} \nwse{e}{a}{\mu} \nwse{e}{b}{\nu}$.
The flat metric $\eta_{ab}$ has $\eta_{-}$ negative eigenvalues and
$\eta_{+} = d - \eta_{-}$ positive eigenvalues.
The space of $p$-forms on $M$ is denoted as $\Omega^{p} \left( M \right)$.

In terms of this metric, we define
$\mathrm{I}^{a_{1} \cdots a_{q}}$
as the map from $p$-forms to $\left(  p-q\right)$-forms,
\begin{equation}
  \mathrm{I}^{a_{1} \cdots a_{q}} :
  \Omega^{p} \left( M \right) \to
  \Omega^{p-q} \left( M \right),
\end{equation}
given by
\begin{equation}
  \mathrm{I}^{a_{1} \cdots a_{q}} =
  \left(
    -1 \right)^{\left( d-p \right) \left( p-q \right) + \eta_{-}}
    \hodge \left( e^{a_{1}} \wedge \cdots \wedge e^{a_{q}} \wedge \hodge
  \right.,
\end{equation}
where $\hodge$ stands for the Hodge dual,
$\hodge : \Omega^{p} \left( M \right) \to \Omega^{d-p} \left( M \right)$,
defined by its action on a $p$-form $\alpha$ as
\begin{equation}
  \hodge \alpha =
  \frac{\sqrt{\left\vert g \right\vert}}{p! \left( d-p \right)!}
  \epsilon_{\mu_{1} \cdots \mu_{p} \nu_{1} \cdots \nu_{d-p}}
  \alpha^{{\mu_{1} \cdots \mu_{p}}}
  \mathrm{d} x^{\nu_{1}} \wedge \cdots \wedge \mathrm{d} x^{\nu_{d-p}}.
\end{equation}

The $\mathrm{I}$-operators have many properties and obey an interesting algebra.
In particular, the most important case is $q=1$,
\begin{equation}
  \mathrm{I}^{a} = \left( -1 \right)^{d \left( p-1 \right) + \eta_{-}}
  \hodge \left( e^{a} \wedge \hodge \right.,
\end{equation}
which satisfies Leibniz's rule,
\begin{equation}
  \mathrm{I}_{a} \left(
    \alpha^{\left( p \right)  } \wedge \beta^{\left( q \right)}
  \right) =
  \mathrm{I}_{a} \alpha^{\left( p \right)} \wedge \beta^{\left( q \right)} +
  \left( -1 \right)^{p}
  \alpha^{\left( p \right)} \wedge \mathrm{I}_{a} \beta^{\left( q\right)}.
\end{equation}


\section{Lorentz Symmetry}
\label{sec:Lorentz}

The antisymmetric piece of $H^{a}$ corresponds to a local Lorentz transformation and can be omitted in all expressions. To show this, let us
consider a generic one-form $H_{a}=H_{ab}e^{b}$ with a symmetric and an antisymmetric piece, $H_{a} = H_{a}^{+} + H_{a}^{-}$, given by
\begin{equation}
  H_{ab}^{\pm} = \frac{1}{2} \left( H_{ab} \pm H_{ba} \right).
\end{equation}

The equation for the first-order piece of $\mathring{u}^{ab}$,
\begin{equation}
  \frac{1}{2} \mathring{\mathrm{D}} H_{a} +
  \mathring{u}_{ab}^{\left( 1 \right)} \wedge e^{b} = 0,
\end{equation}
in this general case takes the form we already know from 
eq.~(\ref{Ec_u_ab(1)})
plus a second term,
\begin{equation}
  \mathring{u}_{ab}^{\left( 1 \right)} = - \frac{1}{2} \left(
    \mathrm{I}_{a} \mathring{\mathrm{D}} H_{b}^{+} -
    \mathrm{I}_{b} \mathring{\mathrm{D}} H_{a}^{+}
  \right)
  -\frac{1}{2} \mathring{\mathrm{D}} H_{ab}^{-}.
  \label{Eq_u_torsionless_antisymmetric}
\end{equation}

Since the perturbation of the vielbein can be written as
\begin{equation}
  \bar{e}_{a} = e_{a} +
  \frac{1}{2} H_{a}^{+} + \frac{1}{2} H_{ab}^{-} e^{b},
  \label{Eq_vielbein_antisymmetric}
\end{equation}
it is straightforward to see that the antisymmetric piece $H_{ab}^{-}$ in eqs.~(\ref{Eq_u_torsionless_antisymmetric}) and~(\ref{Eq_vielbein_antisymmetric}) corresponds to an infinitesimal local Lorentz transformation in the torsionless spin connection and the vielbein,
\begin{align}
 \delta \mathring{\omega}^{ab} & = -\mathring{\mathrm{D}} \lambda^{ab}, &
 \delta e^{a} & = \nwse{\lambda}{a}{b} e^{b},
\end{align}
where $\lambda^{ab} = \frac{1}{2} H_{-}^{ab}$.
Making a slightly different choice for $\Delta^{ab}$ than eq.~(\ref{eq:Delta-constraint}), it is possible to prove explicitly the same for the full geometry with nonvanishing torsion.


\bibliography{IRV_biblio}

\end{document}